\documentclass[showpacs,prb,aps,twocolumn]{revtex4}
\usepackage{graphicx}
\usepackage[normalem]{ulem}

\usepackage{amssymb}

\begin{document}

\title{Vacancy ordering and phonon spectrum in the Fe superconductor
K$_{0.8}$Fe$_{1.6}$Se$_2$}
\author{A. M. Zhang}
\author{K. Liu}
\author{J. H. Xiao}
\author{J. B. He}
\author{D. M. Wang}
\author{G. F. Chen}
\author{B. Normand}
\author{Q. M. Zhang}
\email{qmzhang@ruc.edu.cn} \affiliation{Department of Physics,
Renmin University of China, Beijing 100872, P. R. China}

\date{\today}

\begin{abstract}

We have performed Raman-scattering measurements on a high-quality 
single crystal of the recently discovered Fe-based superconductor 
K$_{0.8}$Fe$_{1.6}$Se$_2$ ($T_c$ = 32 K). At least thirteen phonon modes 
were observed in the wave number range 10$-$300 cm$^{-1}$. The spectra 
possess a four-fold symmetry indicative of bulk vacancy order in the 
Fe-deficient planes. We perform a vibration analysis based on 
first-principles calculations, which both confirms the ordered 
structure and allows a complete mode assignment. We observe an
anomaly at $T_c$ in the 180 cm$^{-1}$ $A_g$ mode, which indicates
a rather specific type of electron-phonon coupling.

\end{abstract}

\pacs{74.70.-b, 74.25.Kc, 63.20.kd, 78.30.-j}

\maketitle

\section{Introduction}

Since the discovery of high-$T_c$ Fe-based superconductors almost
three years ago, much effort has been devoted both to exploring
their physical properties and to the search for new materials in
this class. Until recently, five such systems had been synthesized
and studied, among which Fe(Se,Te) is of particular interest as the 
first Fe-based superconductor not requiring the poisonous element As.
Although the superconducting transition temperature ($T_c$) is low, it
has been found for FeSe that a maximum $T_c$ of 37 K is obtained under a
pressure of 6 GPa.\cite{FeSe37K} This strong dependence has inspired
efforts to find new FeSe-based superconductors by introducing internal
chemical pressure.\cite{MKWu} This idea was realized very recently in
K$_{0.8}$Fe$_2$Se$_2$,\cite{XLChen}  where $T_c \sim 31$ K is close to
that of the ``122'' materials. The system was also identified as being
isostructural with BaFe$_2$As$_2$.\cite{XLChen} Several groups have
now reported similar $T_c$ values by substitution of other alkali metals,
including Rb and Cs.\cite{AFeSe}

The electronic and magnetic properties of the A$_x$Fe$_{2-y}$Se$_2$ compounds 
have been probed extensively. Angle-resolved photoemission spectroscopy 
(ARPES) measurements\cite{ARPES,newARPES} agree on the presence of one 
Fermi surface around the M point, and on a conventional $s$-wave pairing 
symmetry, but differ on further details. Nuclear magnetic resonance (NMR)
measurements in superconducting crystals find very narrow line
widths,\cite{NMR} strong coupling, singlet superconductivity but with no 
coherence peak, and very weak spin fluctuations.\cite{NMR,NMR2,NMRorder} 
Infrared optical conductivity measurements have identified the parent 
compound as a small-gap semiconductor and not a Mott insulator.\cite{Infrared}
Neutron diffraction studies find a very large ordered moment and high magnetic 
transition temperature.\cite{wbnd} High-pressure experiments by several groups
have led to contradictions which indicate extreme sensitivity to the
nature of the defects.\cite{Pressure}

These high-pressure measurements are not the only indications for a crucial
role of Fe vacancies. Superconductivity itself occurs only in Fe-deficient
samples. A broad resistivity peak around 150 K is reduced dramatically by
altering the Fe content.\cite{GFChen} The unexpected abundance of
infrared-active phonon modes\cite{Infrared} may also be attributed to
Fe vacancies. Resolving the distribution of Fe vacancies on the microscopic
level is clearly essential for a full understanding of the electronic and
magnetic properties. A study exploiting the unique sensitivity of Raman
scattering to probe local lattice symmetries is thus required.

Here we measure Raman-scattering spectra in a high-quality superconducting 
crystal of K$_{0.8}$Fe$_{1.6}$Se$_2$, as described in Sec.~II. Our results, 
presented in Sec.~III, contain at least thirteen phonon modes, far more 
than would be expected for a normal 122 structure. The spectra we obtain 
by rotating the crystal show a four-fold symmetry, which we identify as 
C$_{4h}$ or C$_4$. These results indicate an ordered structure of Fe vacancies, 
and in Sec.~IV we assume the highest-symmetry defect configuration to perform 
a first-principles vibration analysis. The calculated mode frequencies and 
symmetry assignments are fully consistent with our data, confirming completely 
the vacancy ordering pattern. Our temperature-dependent spectra, shown in 
Sec.~V, reveal one specific $A_g$ mode with a clear frequency jump at $T_c$, 
which suggests a close connection between phonons and superconductivity. 
Section VI contains a brief summary. 

\begin{figure}[t!]
\includegraphics[width=8.2cm,angle=0]{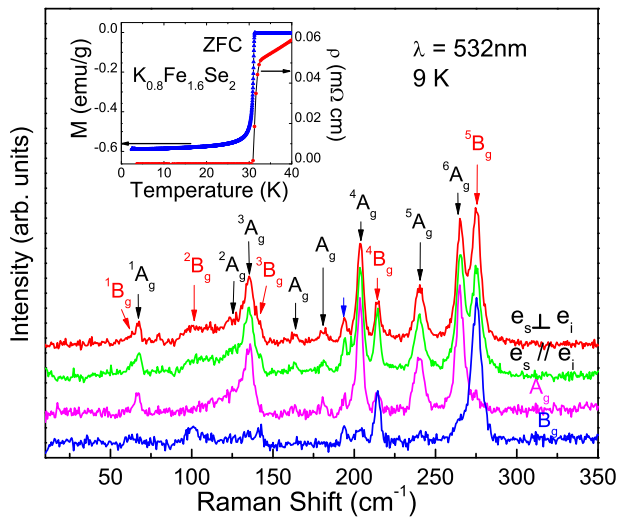}
\caption{(Color online) Raman spectra of a K$_{0.8}$Fe$_{1.6}$Se$_2$ crystal
at 9 K. Labels $e_i$ and $e_s$ denote the incident and scattered light
polarizations. Black and red arrows indicate respectively the assigned
$A_g$ and $B_g$ modes, the blue arrow a weak mode whose symmetry remains
unknown. The accumulation time for the cross-polarization measurements is
approximately three times that in parallel polarization. The spectra labeled
$A_g$ and $B_g$ were measured by controlling the relative angle between
the incident and scattered light polarizations once the mode symmetries
had been assigned; this assignment was performed by combining rotation
measurements (Fig.~2) with the formal symmetry analysis [Eq.~(1)]. Inset:
resistivity and magnetization data for the same crystal.}
\label{fig1}
\end{figure}

\section{Material and Method}

The crystal was grown by the Bridgeman method. The detailed growth
procedure may be found elsewhere.\cite{GFChen} The most accurate
available composition determination is the neutron diffraction
refinement,\cite{wbnd} which gives K$_{0.8}$Fe$_{1.6}$Se$_2$.
X-ray diffraction patterns indicate no discernible secondary phase.
The resistivity was measured by a Quantum Design PPMS and the
magnetization by the PPMS vibrating sample magnetometer (VSM): both
quantities, shown in the inset of Fig.~1, exhibit transitions at
$T_c \sim$ 32 K, the diamagnetic transition appearing particularly
sharp. Although the precise demagnetization factor is not available,
the estimated superconducting volume fraction is (from the sharp
diamagnetism and absence of a secondary phase) close to 100\%.
These results indicate that the crystal is of very high quality.

The crystal was cleaved in a glove box and a flat, shiny surface was
obtained. The surface is parallel to the ($ab$)-plane, and forms the
basal plane for our Raman measurements. The two pieces of freshly-cleaved
crystal were sealed under an argon atmosphere: one was transferred into the
cryostat within 30 seconds for Raman-scattering studies and the other used
for symmetry investigations by rotating the crystal. The cryostat was
evacuated immediately to a work vacuum of approximately $10^{-8}$ mbar.
Raman measurements were performed in a pseudo-backscattering configuration
with a triple-grating monochromator (Jobin Yvon T64000), delivering a
spectral resolution better than 0.6 cm$^{-1}$. The light source is a 532 nm
solid-state laser (Torus 532, Laser Quantum) whose beam was focused into
a spot of diameter {\it ca.}~10--20 $\mu$m on the sample surface. The beam
power is lower than 1 mW, and the real temperature at the spot was calibrated
using the intensity relation between the Stokes and anti-Stokes spectra.

\begin{figure}[t!]
\includegraphics[width=8.2cm,angle=0]{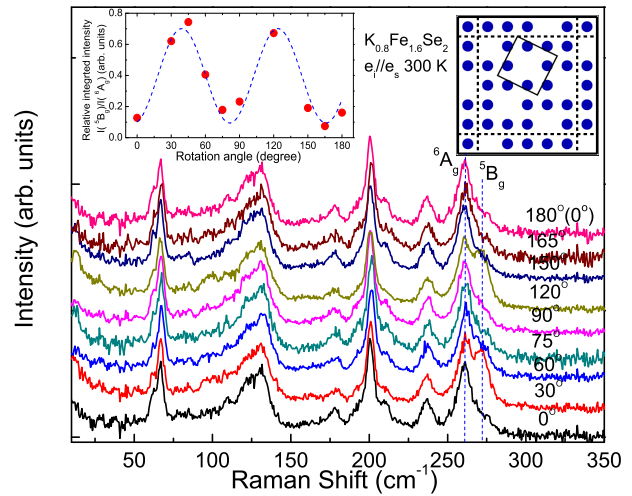}
\caption{(Color online) Raman spectra obtained by rotation, collected in
air at room temperature immediately after cleavage. The zero-degree point
is chosen arbitrarily (not measured relative to the crystal principal axes).
Dashed lines mark two modes, labeled as $^5B_g$ and $^6A_g$, whose relative
integrated intensity variation is shown as a function of angle in the left
inset [the dashed line has the form $\sin^22\theta$].  Right inset:
schematic representation of ordered vacancy configuration with 20\% Fe
deficiency and four-fold symmetry.} \label{fig2}
\end{figure}

\section{Phonon Raman Spectra: Symmetry and Vacancy Order}

The phonon modes of K$_{0.8}$Fe$_{1.6}$Se$_2$ are shown in the
low-temperature spectra of Fig.~1. The most striking feature is that
at least thirteen modes are present in the spectra collected with both
parallel and cross-polarization configurations. All of these modes are
located below 300 cm$^{-1}$ (Fig.~4 illustrates this most clearly). Our
results are therefore consistent both by number and location with the
infrared optical spectra.\cite{Infrared} We stress that only four
Raman-active modes are present in 122,\cite{Iliev} to which
K$_{0.8}$Fe$_{1.6}$Se$_2$ was thought to be isostructural, and that 
the spectra of Fig.~1 are completely different from those observed in 
Fe(Se,Te).\cite{QMZhang} The surprisingly large number of Raman and 
infrared phonon modes reinforces the assumption that the Fe deficiency drives
a local breaking of lattice symmetry. A vibration analysis of the modes
in Fig.~1, which we perform below, is thus significantly more complicated
than for 122 and requires additional structural information.

In Fig.~2 we present spectra obtained by rotating the crystal. These
display clearly a four-fold symmetry, which is highlighted in the right
inset. A formal analysis based on four-fold symmetry proceeds as follows.
There are two possible Raman tensors, one for $C_{4h}$/$C_4$ and the other
for $D_{4h}$/$D_4$.\cite{Hayes} If $\theta$ is defined as the angle between
the crystallographic $a$-axis and the polarization of the incident light,
and $\varphi$ as the angle between the polarizations of the incident and
scattered light, then the angle-dependences of the Raman intensities for
$C_{4h}$/$C_4$ take the form
\begin{eqnarray}
I_{A_{g}} & = & |a \cos(\varphi) + c \sin(\varphi)|^{2}, \nonumber \\
I_{B_{g}} & = & |d\cos(\varphi + 2\theta) + e\sin(\varphi + 2\theta)|^{2},
\end{eqnarray}
where $a$, $c$, $d$, and $e$ are matrix elements of the Raman tensors, and
depend on linear combinations of polarizability derivatives for the atoms
involved in each mode. Similarly, for $D_{4h}$/$D_4$ one has
\begin{equation}
I_{A_{1g}}  = |a \cos(\varphi)|^{2}, \;\;\;\;
I_{B_{1g}} \; = \; |e\sin(\varphi + 2\theta)|^{2}.
\end{equation}
Because rotating the crystal varies the angle $\theta$, it is clear that
modes with constant intensities are all $A$-type ($A_{1g}$, $A_g$, $A_1$, or
$A$), while those showing an intensity modulation are $B$-type. Further,
Eq.~(2) prohibits the observation of $A$-type modes in $D_{4h}$ symmetry
in a cross-polarized configuration ($\varphi = \pi/2$), which contradicts
the observations in Fig.~1. Thus our results exclude explicitly the D$_{4h}$
symmetry expected for the 122 structure, and allow us to deduce that the
crystal symmetry is C$_{4h}$ or C$_4$. The $A$-type modes in Fig.~1 may
therefore be identified as $A_g$ or $A$ phonons. This symmetry constrains
very strongly the possible vacancy ordering patterns.

\begin{figure}[t!]
\includegraphics[width=7cm,angle=0]{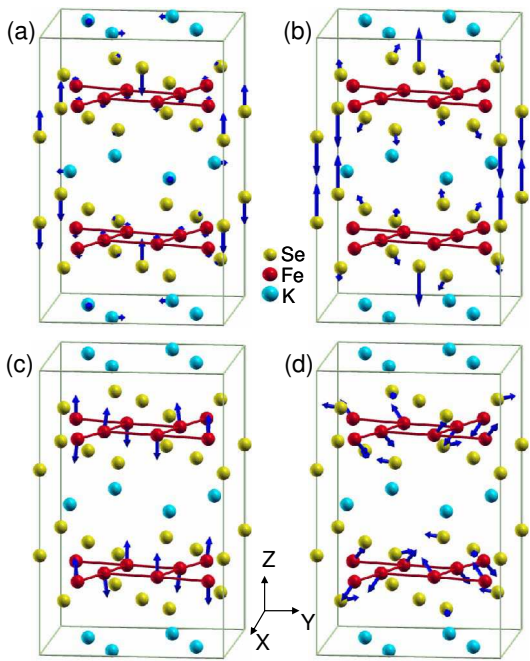}
\caption{(Color online) Displacement patterns for selected Raman-active
$A_{g}$ [(a) and (b), labeled as $^1A_g$ and $^4A_g$ in Table I] and $B_{g}$
modes [(c) and (d), labeled as $^4B_g$ and $^5B_g$ in Table I] of
K$_{0.8}$Fe$_{1.6}$Se$_{2}$. Fe atoms connected by red lines have
right-handed chirality. Amplitude arrows in (d) are scaled up by a factor
of 1.4 over the other panels. Atomic structures and displacement patterns
prepared using XCRYSDEN.\cite{kokalj}}
\label{figf3}
\end{figure}

Indications for vacancy ordering in A$_x$Fe$_{2-y}$Se$_2$ systems
have been obtained in several recent measurements. Imaging by
transmission electron microscopy (TEM)\cite{rTEM} reveals rather
well-defined surface order for $y = 0.5$ and $y = 0.2$. NMR
measurements\cite{NMRorder} find additional line-splittings. While
Raman scattering is a surface technique, the wavelength, laser spot
size (10-20 $\mu$m), and penetration depth mean that macroscopic
symmetries are probed. The 20\% vacancy concentration in our crystal
($y = 0.4$) suggests the structure illustrated in the inset of
Fig.~2: this arrangement has a $\sqrt{5}$$\times$$\sqrt{5}$ unit
cell, has local four-fold symmetry, and, crucially, has lost the two
mirror planes perpendicular to the $(ab)$-plane, in complete
consistency with our Raman measurements. This microscopic structure
has been demonstrated spectacularly by the very recent neutron
diffraction studies of Ref.~\onlinecite{wbnd}. X-ray diffraction 
studies of several crystals with $y = 0.37$--0.39 have confirmed 
this result.\cite{rXRD}

\section{First-Principles Calculations}

To compare with the observed phonon modes, we have calculated the nonmagnetic
electronic structure and zone-center phonons of K$_{0.8}$Fe$_{1.6}$Se$_{2}$
from first-principles density functional calculations. We use the Vienna
{\it ab-initio} simulation package,\cite{kresse} which is based on the
projector augmented wave (PAW) method\cite{paw} with a general gradient
approximation (GGA)\cite{pbe} for exchange-correlation potentials. The
structural information determined by neutron diffraction\cite{wbnd} is
that K$_{0.8}$Fe$_{1.6}$Se$_{2}$ has space group I4/m and point group
$C_{4h}$. The Wyckoff positions of the K, Fe, and Se atoms and the
corresponding symmetry analysis are listed in Table I. The frequencies
and displacement patterns of the phonon modes were calculated using the
dynamical matrix method.\cite{liu05}

\begin{table}[!b]
\caption{Symmetry analysis for I4/m and assignment of Raman-active vibration
modes in K$_{0.8}$Fe$_{1.6}$Se$_{2}$}
\begin{center}
\begin{tabular*}{8.4cm}{@{\extracolsep{\fill}} cccccccc}
\hline \hline
Atom & Wyckoff & \multicolumn{2}{c}{Optical modes} & \\
\cline{3-4}
 & position & Raman active & Infrared active & \\
\hline
K  & 8h & $2A_{g}+2B_{g}+2E_{g}$ & $A_{u}+4E_{u}$ & \\
Fe & 16i & $3A_{g}+3B_{g}+6E_{g}$ & $3A_{u}+6E_{u}$ & \\
Se & 4e & $A_{g}+2E_{g}$ & $A_{u}+2E_{u}$ & \\
Se & 16i & $3A_{g}+3B_{g}+6E_{g}$ & $3A_{u}+6E_{u}$ & \\
\hline \hline
Expt. Freq. & Cal. Freq. & \multicolumn{2}{c}{Symmetry~~~~~Atoms~~~~~Index} \\
(cm$^{-1}$) & (cm$^{-1}$) & & \\
\hline
61.4 & 66.7 & \multicolumn{2}{c}{~~~~$B_{g}$~~~~~~~~~~~~~Se~~~~~~~~~~$^1B_{g}$} \\
66.3 & 75.1 & \multicolumn{2}{c}{~~~~$A_{g}$~~~~~~~~~~~~~Se~~~~~~~~~~$^1A_{g}$} \\
100.6 & 106.2 & \multicolumn{2}{c}{~~~~$B_{g}$~~~~~~~~~~~~~K~~~~~~~~~~$^2B_{g}$} \\
123.8 & 130.5 & \multicolumn{2}{c}{~~~~$A_{g}$~~~~~~~~~~~~~Se~~~~~~~~~$^2A_{g}$} \\
134.6 & 159.2 & \multicolumn{2}{c}{~~~~$A_{g}$~~~~~~~~~~~~~Se~~~~~~~~~$^3A_{g}$} \\
141.7 & 149.0 & \multicolumn{2}{c}{~~~~$B_{g}$~~~~~~~~~~~~~Se~~~~~~~~~$^3B_{g}$} \\
202.9 & 212.6 & \multicolumn{2}{c}{~~~~$A_{g}$~~~~~~~~~~~~~Se~~~~~~~~~$^4A_{g}$} \\
214.3 & 238.3 & \multicolumn{2}{c}{~~~~$B_{g}$~~~~~~~~~~~~~Fe~~~~~~~~~$^4B_{g}$} \\
239.4 & 268.5 & \multicolumn{2}{c}{~~~~$A_{g}$~~~~~~~~~~~~~Fe~~~~~~~~~$^5A_{g}$} \\
264.6 & 286.1 & \multicolumn{2}{c}{~~~~$A_{g}$~~~~~~~~~~~~Fe,Se~~~~~~~$^6A_{g}$} \\
274.9 & 279.0 & \multicolumn{2}{c}{~~~~$B_{g}$~~~~~~~~~~~~~Fe~~~~~~~~~$^5B_{g}$} \\
\hline \hline
\end{tabular*}
\end{center}
\end{table}

In total, 17 $A_g$ and $B_g$ modes are allowed by the crystal symmetry,
among which 11 of the modes in the polarized Raman spectra of Fig.~1 may
be assigned accurately from the calculations (Table I). The vibration
patterns of selected modes are shown in Fig.~3. The calculated phonon
frequencies, 75.1 and 212.6 cm$^{-1}$ for $A_g$ modes [Fig.~3(a) and (b)]
and 238.3 and 279.0 cm$^{-1}$ for $B_g$ modes [Fig.~3(c) and (d)], agree
very well with the 66.3, 202.9, 214.3, and 274.9 cm$^{-1}$ modes in the
experimental measurements. The $A_g$ and $B_g$ modes shown in Figs.~3(b)
and (c) correspond exactly to the $A_{1g}$ mode of As and $B_{1g}$ mode
of Fe in the 122 system.\cite{Iliev} The success of the phonon assignment
confirms completely the assumed pattern of vacancy ordering. Only three of
the observed 13 modes cannot be assigned well, even though two have $A_g$
character. We suggest that these may originate from distortions of the
Fe-Se layer, which lie outside the symmetry constraints applied.

\begin{figure}[t!]
\includegraphics[bb=0 0 820 553,width=8cm,height=5.6cm,angle=0]{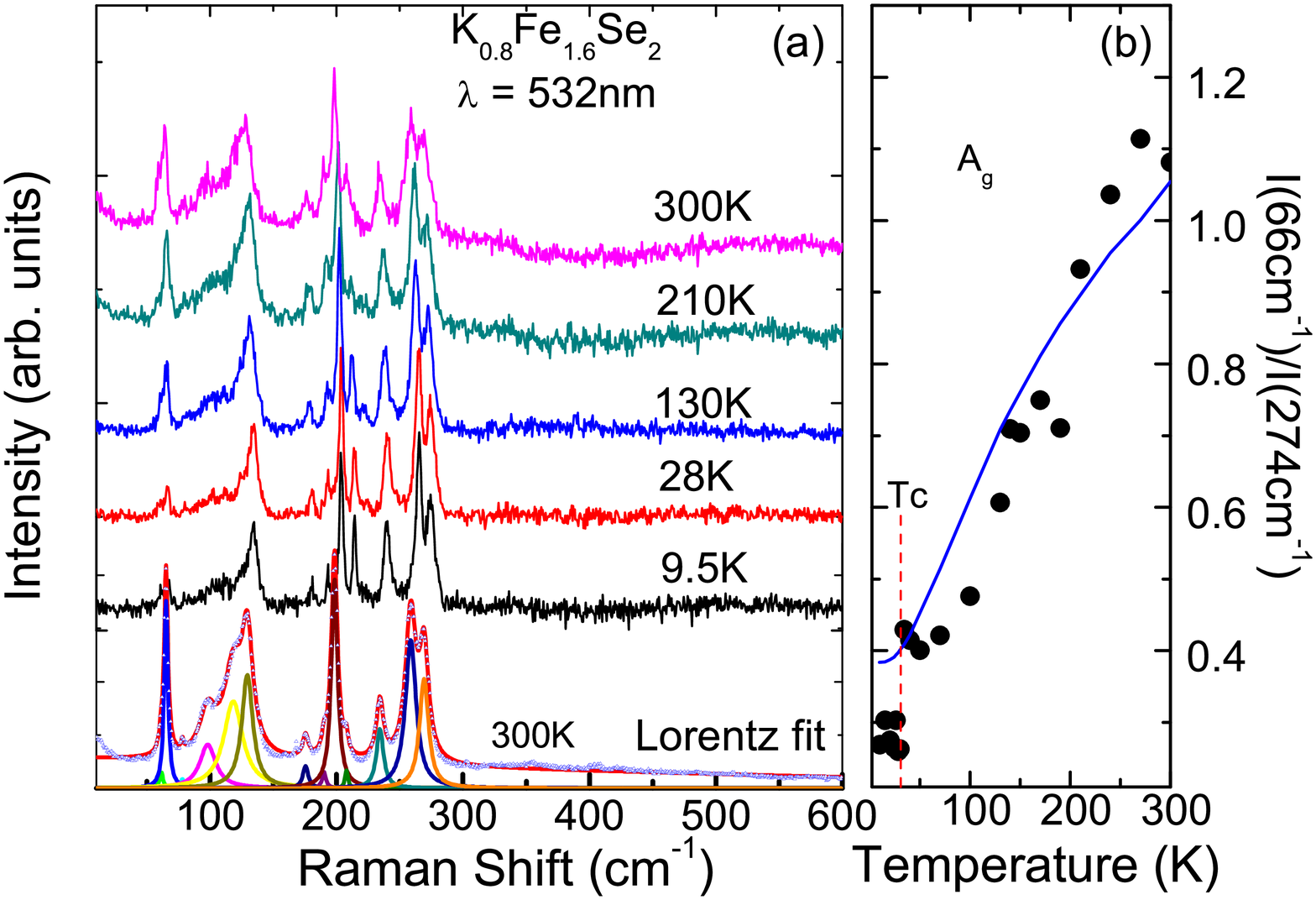}
\caption{(Color online) (a) Raman spectra at selected temperatures.
An example of fitting with Lorentz functions is shown at the base of the
panel for 300 K data collected in parallel polarization with an accumulation
time four times that of the other spectra. The frequency resolution is then
0.3 cm$^{-1}$, which allows the explicit separation of 13 modes in the narrow
frequency interval 50--300 cm$^{-1}$. The fits used to obtain the data points
in Figs.~4(b) and 5 have a coefficient of determination $r^2 > 0.95$. (b)
Temperature-dependence of the relative intensities of the 66 cm$^{-1}$ and
274 cm$^{-1}$ modes. The solid line shows the Bose-Einstein thermal factor.}
\label{fig4}
\end{figure}

\section{Temperature-Dependent Spectra}

We have also measured Raman spectra over the full temperature range to
300 K [Fig.~4(a)]. The temperature-dependence of the phonon modes is
described explicitly by fitting with Lorentz functions. All relative
intensities show a conventional phonon form, with one exception (below).
This line-shape analysis shows no evidence of the Fano asymmetry
characteristic of electron-phonon coupling. However, all of the phonons are
rather broad when compared with those in 122 systems,\cite{Iliev} due perhaps
to intrinsic disorder (incompletely ordered vacancies). This analysis alone
cannot therefore exclude a generic electron-phonon coupling in
K$_{0.8}$Fe$_{1.6}$Se$_2$.

The $A_g$ mode at 66 cm$^{-1}$ behaves rather anomalously. The steep fall 
in its relative intensity [Fig.~4(b)] follows a Bose-Einstein form only 
to lowest order, but with significant deviations. In addition to a clear 
intensity jump at $T_c$, there is a further apparent anomaly around 160 K, 
where X-ray diffraction studies find no structural transition.\cite{Pressure} 
We have confirmed (Table I) that the 66 cm$^{-1}$ phonon is the vibration mode 
of a Se atom, and suggest that this is the origin of its unusual behavior.

The temperature-dependence of selected phonon frequencies is shown
in Fig.~5. The superconducting transition has very little effect on
the frequency in every case but one: the $A_g$ mode at 180 cm$^{-1}$
exhibits a clear jump of approximately 1 cm$^{-1}$ at $T_c$
[Fig.~5(a)]. This anomaly is definite evidence for a particular type
of connection between phonons and superconductivity, and it
constitutes a statement concerning both symmetry and energies. For
the former, the symmetry relation between the superconducting gaps
on the available Fermi surfaces should be consistent with $A_g$. For
the latter, we suggest that a coupling of this phonon mode to the
superconducting quasiparticles may be visible\cite{rslin} in ARPES
spectra, where it should be expected at energies $E \sim $20$-$25
meV and at the $\Gamma$ point (the part of the Brillouin zone probed
by Raman scattering). Anomalous features are indeed present in this
region in the two most recent ARPES studies of A$_x$Fe$_{2-y}$Se$_2$
superconductors,\cite{newARPES} and we propose that high-resolution
ARPES measurements may reveal the physics behind this phenomenon.

\begin{figure}[t]
\includegraphics[width=8.5cm,angle=0]{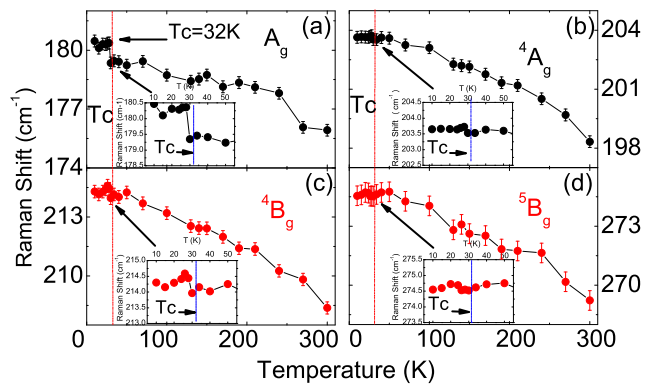}
\caption{(Color online) (a)-(d) Temperature-dependence of phonon
frequencies for four modes with the highest measured intensities.
Insets magnify the region around $T_c$. Horizontal and vertical
scales are the same in all panels and insets.} \label{fig5}
\end{figure}

\section{Summary}

To conclude, we have performed Raman-scattering measurements on the
recently discovered Fe-based superconductor K$_{0.8}$Fe$_{1.6}$Se$_2$.
Using a high-quality single crystal, we find that there exist at
least thirteen phonon modes and that the spectra have four-fold
symmetry. This demonstrates the presence of long-range-ordered
configurations of Fe vacancies, which dictate the local lattice
vibrations. We perform first-principles calculations to obtain a
complete and consistent phonon assignment, confirming the nature of
the vacancy ordering pattern. We find that only one mode exhibits a
change in frequency around $T_c$, suggesting a rather specific
connection between superconductivity and lattice vibrations.

\acknowledgments

We thank W. Bao and Z. Y. Lu for helpful discussions. This work was
supported by the 973 program under Grant No.~2011CBA00112, by the
NSF of China under Grant Nos.~11034012 and 11004243, by the
Fundamental Research Funds for Central Universities, and by the
Research Funds of Renmin University.

\end{document}